\documentclass[aps,showpacs,superscriptaddress,groupedaddress, twocolumn]{revtex4-1} 

\pdfoutput=1

\usepackage{amsmath}
\usepackage{amssymb}

\usepackage{graphicx}
\usepackage{subfigure}
\usepackage{longtable}

\usepackage{dcolumn}
\usepackage{bm}
\usepackage{array}
\usepackage{color}

\usepackage{longtable}

\graphicspath{{../1figures/}}

\begin{document}

\title{Controllability and maximum matchings of complex networks}

\author{Jin-Hua Zhao$^1$}
 \email{Corresponding author: zhaojh190@gmail.com}
 
\author{Hai-Jun Zhou$^{1, 2}$}

\affiliation{$^1$
CAS Key Laboratory of Theoretical Physics,
Institute of Theoretical Physics,
Chinese Academy of Sciences,
Beijing 100190,
China}

\affiliation{$^2$
School of Physical Sciences,
University of Chinese Academy of Sciences,
Beijing 100049,
China}

\date{\today}

\begin{abstract}
Previously,
the controllability problem of a linear time-invariant dynamical system
was mapped to the maximum matching (MM) problem
on the bipartite representation of the underlying directed graph,
and the sizes of MMs on random bipartite graphs
were calculated analytically with the cavity method at zero temperature limit.
Here we present an alternative theory to estimate MM sizes
based on the core percolation theory and the perfect matching of cores.
Our theory is much more simplified and easily interpreted,
and can estimate MM sizes on random graphs
with or without symmetry between out- and in-degree distributions.
Our result helps to illuminate the fundamental connection
between the controllability problem and the underlying structure of complex systems.
\end{abstract}

\maketitle



\section{Introduction}

Graph theory and network science
\cite{
Bollobas-1998,
Albert.Barabasi-RMP-2002,
Newman-SIAM-2003,
Boccaleti.Latora.Moreno.Chavez.Hwang-PhysRep-2006,
Dorogovtsev.Goltsev-RMP-2008,
Newman-2018}
provide a consistent framework to understand the structure and the dynamics of complex connected systems.
In the past few years,
the problems of controllability and control on complex networks
and their application on real-world networked systems
are among the main focuses of complex systems community
\cite{
Kalman-JSocIndusApplMath-1963,
Lin-IEEE-1974,
Liu.Slotine.Barabasi-Nature-2011,
Liu.Slotine.Barabari-PLoSONE-2012,
Posfai.etal-SciRep-2013,
Jia.etal-NatCommun-2013,
Cornelius.Kath.Motter-NatCommun-2013,
Jia.Posfai-SciRep-2014,
Menichetti.DallAsta.Bianconi-PRL-2014,
Nepusz.Vicsek-NatPhys-2012,
Yan.etal-PRL-2012,
Yan.etal-NatPhys-2015,
Wang.etal-NatCommun-2013,
Gao.etal-NatCommun-2014,
Sun.Motter-PRL-2013,
Ruths.Ruths-Science-2014,
Li.etal-Science-2017,
Vinayagam.etal-PNAS-2016,
Yan.etal-Nature-2017}.
A rather comprehensive review on control issues in complex systems can be found in
\cite{
Liu.Barabasi-RMP-2016}.
Here we focus on a controllability problem that has been frequently studied in the recent network community,
the controllability of linear-time invariant systems with nodal dynamics
\cite{
Kalman-JSocIndusApplMath-1963,
Lin-IEEE-1974}.
In the paper
\cite{
Liu.Slotine.Barabasi-Nature-2011},
the problem of finding the minimal driver node sets (MDNSs) or the minimal actuator sets
to steer system dynamics
is mapped to the maximum matching (MM) problem
\cite{
Lovasz.Plummer-1986}
(finding a maximal set of edges with no shared vertices)
on the bipartite representation of the underlying directed graph.
On the analytical side,
the cavity method at the zero temperature limit
is derived to estimate MM sizes
(size of a set of matched edges as a MM solution)
\cite{
Liu.Slotine.Barabasi-Nature-2011,
Menichetti.DallAsta.Bianconi-PRL-2014,
Mezard.Montanari-2009,
Zhou.OuYang-arXiv-2003,
Zdeborova.Mezard-JStatPhys-2006,
Kreacic.Bianconi-arXiv-2018}.
On the simulation side,
the Hopcroft-Karp algorithm
 \cite{
 Hopcroft.Karp-SIAMJComput-1973}
is adopted to find the MM solutions.
Yet the analytical framework presented in
\cite{
Liu.Slotine.Barabasi-Nature-2011}
suffers from two points.
First,
the final equation connecting graph structure (the degree distribution)
and MM sizes on random graphs [Eq. (S37)] is explicit,
yet its derivation is quite complicated,
burying a potentially simple and intrinsic explanation of the problem.
Second,
the degree asymmetry,
or the disparity between out- and in-degree distributions,
which is ubiquitous in real-world networked systems,
is still not well fit in the above analytical framework,
even its implication in controllability and MM sizes has been touched on in some of the literature
\cite{
Liu.Slotine.Barabasi-Nature-2011,
Jia.etal-NatCommun-2013,
Jia.Posfai-SciRep-2014}.

In this paper,
we try to resolve the above two points in a single framework.
Our main contribution
is the derivation of an alternative analytical framework
to estimate MM sizes on general random graphs
based on the intuition of the Karp-Sipser algorithm
\cite{
Karp.Sipser-IEEFoCS-1981,
Aronson.Frieze.Pittel-RandStrucAlgo-1998}
and the related core percolation theory
\cite{
Stauffer.Aharony-1994,
Bauer.Golinelli-EPJB-2001,
Liu.Csoka.Zhou.Posfai-PRL-2012,
Zhao.Zhou-arXiv-2018}.
Our theory is much more simplified and easily interpreted.
On random graphs with degree symmetry,
our theory simply retrieves the result for MDNS sizes derived with the cavity method at the zero temperature limit
in \cite{
Liu.Slotine.Barabasi-Nature-2011}.
On random graphs without degree symmetry,
our theory works naturally,
thus provides an analytical perspective to the implication of degree asymmetry in the network controllability problem.
In all, our framework helps to clarify a theoretical understanding of the roles of network topology and structure
in the controllability of complex connected systems.


\section{Model}
\label{sec:model}

%
%
\begin{figure}
\begin{center}
 \includegraphics[width = 0.99 \linewidth]{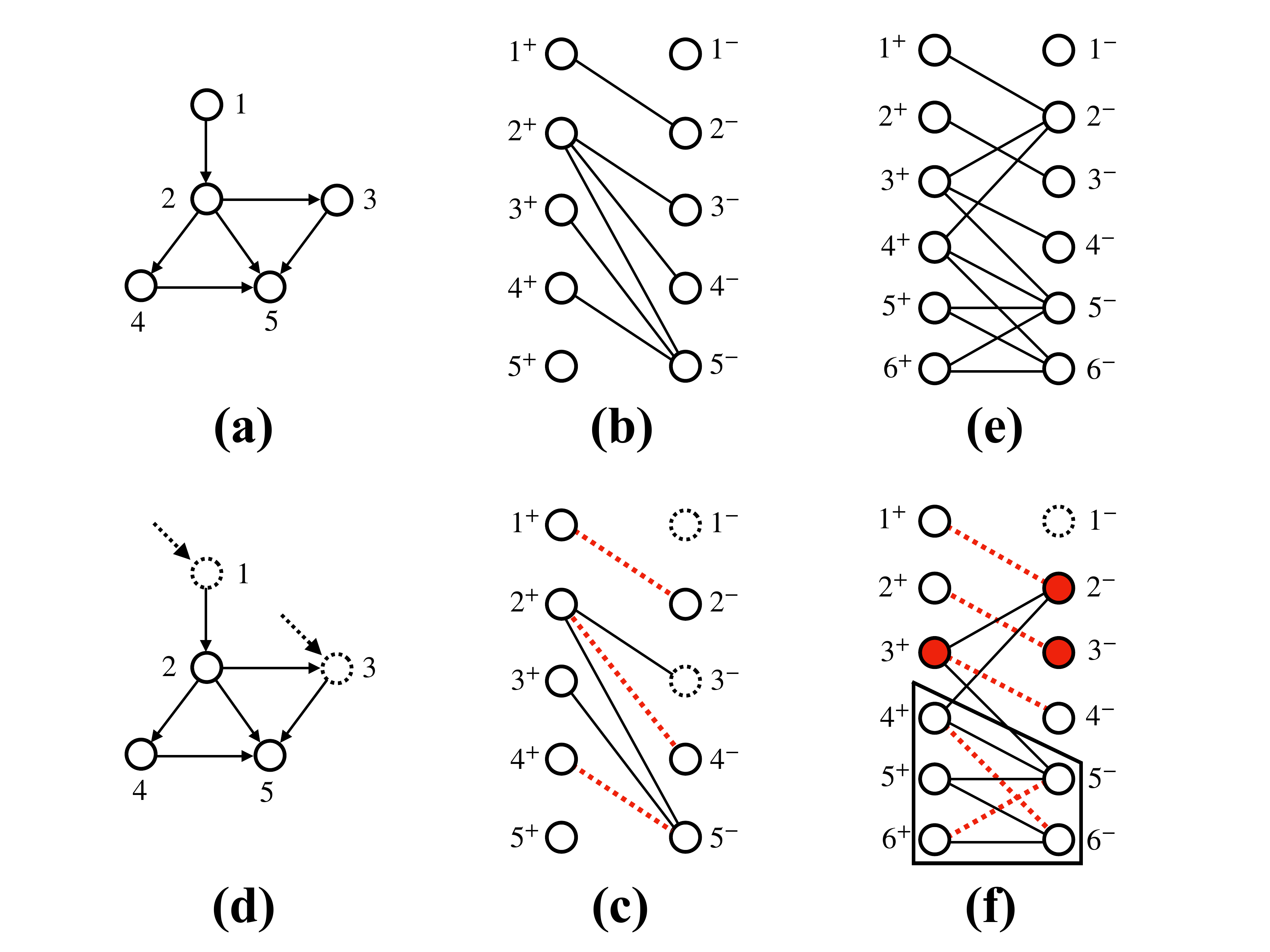} 
\end{center}
\caption{
 \label{fig:model}
An example of a directed graph,
its bipartite representation,
its MM and MDNS,
and the Karp-Sipser algorithm on it.
(a) A small directed graph with five vertices and six arcs.
(b) The undirected bipartite representation of (a).
(c) A MM of three edges in red dashed lines on the undirected graph in (b).
Correspondingly,
the in-vertices in dashed circles are unmatched.
(d) A configuration of MDNS,
in which the driver nodes or the actuators in dashed circles are simply the unmatched in-vertices in (c).
(e) A bipartite representation of a hypothetical directed graph
with six vertices and $12$ arcs.
(f) A MM of five edges in red dashed lines
from the Karp-Sipser algorithm on the undirected graph in (e).
The first stage of the GLR procedure consists of three elementary steps,
after which the three red filled circles outside the trapezoid are roots
and the five vertices and the $6$ edges enclosed in the trapezoid form the core.
The three dashed edges adjacent to these roots
are matched edges constructed from the GLR procedure.
In the core, a randomized edge removal procedure is applied
as an edge $(4^{+}, 6^{-})$ is removed and correspondingly matched.
On the residual graph after the randomized edge removal procedure,
the second stage of the GLR procedure involves only one elementary step,
in which the root $5^{-}$ is removed.
Correspondingly, an edge $(6^{+}, 5^{-})$
[chosen randomly from $(5^{+}, 5^{-})$ and $(6^{+}, 5^{-})$] is matched.
Till now the bipartite graph (f) leaves no edge
and the set of dashed edges constitutes a matching for (e).
The in-vertex $1^{-}$ in a dashed circle is the only unmatched vertex in the in-vertex set.}
\end{figure}
%
%

We explain here the controllability and the MM problems on directed graphs.
A directed graph $D = \{V, E\}$
consists of a vertex set $V$ of a size $N (= |V|)$
and an arc set $E$ of a size $M (= |E|)$ among vertices.
An ordered pair $\{i, j\}$ of vertices $i$ and $j$
denotes an arc (a directed edge) starting from $i$ and arriving at $j$.
The undirected bipartite representation $\Gamma$ of $D$ is derived as follows:
each vertex $i \in V$ is split into an out-vertex $i^{+}$ and an in-vertex $i^{-}$;
each arc $\{i, j\} \in E$ leads to an undirected edge $(i^{+}, j^{-})$ as an unordered pair of vertices;
thus $\Gamma = \{V^{+} \cup V^{-}, E_{-}^{+}\}$,
in which $V^{+}$, $V^{-}$, and $E_{-}^{+}$
are the out-vertex set, the in-vertex set, and the undirected edge set constructed as above, respectively.
Once we find a MM of the undirected bipartite graph $\Gamma$,
the unmatched vertices in the in-vertex set constitute a MDNS
\cite{
Liu.Slotine.Barabasi-Nature-2011}.
Specifically, $N_{\rm{D}} = \max \{1, N - N_{\rm{MM}}\}$
in which $N_{\rm{MM}}$ is the size of a MM
and $N_{\rm{D}}$ is the size of the corresponding MDNS.
On large graphs,
we have $n_{\rm{D}} = 1 - y$
in which $y (\equiv N_{\rm{MM}} / N)$ is the relative size or the fraction of a MM
and $n_{\rm{D}} (\equiv N_{\rm{D}} / N)$ is the relative size or the fraction of the corresponding MDNS.
An illustration of the above notions is in Figs. \ref{fig:model} (a) to \ref{fig:model} (d).

On the simulation method to find MM solutions on undirected bipartite graphs,
instead of the usually adopted Hopcroft-Karp algorithm
\cite{
Hopcroft.Karp-SIAMJComput-1973},
we adopt the Karp-Sipser algorithm
\cite{
Karp.Sipser-IEEFoCS-1981},
which is basically a randomized local algorithm
and finds MMs on a given undirected graph with high probability.
A basic component of the Karp-Sipser algorithm
is the greedy leaf removal (GLR) procedure,
in which iteratively any vertex with only one nearest neighbor (a leaf)
is removed along with its sole nearest neighbor (a root)
in an elementary step.
The GLR procedure
leads to the core percolation on graphs
\cite{
Bauer.Golinelli-EPJB-2001,
Liu.Csoka.Zhou.Posfai-PRL-2012},
and its original form and variants have implications
in various combinatorial optimization and satisfiability problems
\cite{
Karp.Sipser-IEEFoCS-1981,
Aronson.Frieze.Pittel-RandStrucAlgo-1998,
Zhou.OuYang-arXiv-2003,
Zdeborova.Mezard-JStatPhys-2006,
Zhao.Zhou-arXiv-2018,
Weigt.Hartmann-PRL-2000,
Mezard.RicciTersenghi.Zecchina-JStatPhys-2003,
Cocco.etal-PRL-2003,
Correale.etal-PRL-2006,
Lucibello.RicciTersenghi-IntJStatMech-2014,
Zhao.Zhou-JStatPhys-2015,
Zhao.Zhou-LNCS-2015}.
The Karp-Sipser algorithm,
an iterative process of the GLR procedures and the randomized edge removal procedures,
is basically both a graph pruning process with vertex and edge removals
and a local algorithm to construct a matching.
To construct a matching on a given graph,
once an edge is matched and added into a matching,
all the adjacent edges to its end-vertices are removed since,
as the local constraint,
there is at most one matched edge at which a vertex can be adjacent.
On a general graph,
the Karp-Sipser algorithm involves multiple stages of the GLR procedure,
while in each stage the GLR procedure
(with multiple elementary steps of removing a leaf and its neighboring root)
is applied on the current graph to reveal its core.
Correspondingly,
in each elementary step of the GLR procedure,
the edge between a root and its sole leaf
or between a root and one of its multiple neighboring leaves
is matched and added into a matching
\cite{
Karp.Sipser-IEEFoCS-1981,
Bauer.Golinelli-EPJB-2001}.
On the core of the current graph, the randomized edge removal procedure is further applied,
in which an edge in the core is randomly chosen and removed
along with all the adjacent edges of the two end-vertices.
Correspondingly, the removed edge is matched and added into a matching.
The removal of an edge in a core possibly leads to newly generated leaves in the residual graph,
thus triggers a new stage of the GLR procedure.
Thus upon the iterative Karp-Sipser algorithm,
a given graph shrinks to a graph without edges,
correspondingly the matched edges constructed from the GLR procedures
and the randomized edge removal procedures constitute an approximate MM of the given graph.
An example of the Karp-Sipser algorithm is in Figs. \ref{fig:model} (e) and \ref{fig:model} (f).

\section{Theory}
\label{sec:theory}

Here we derive an analytical framework to estimate
the fraction of MMs on random undirected bipartite graphs.
Before presenting our theory,
we explain some graphical notions.
On an undirected bipartite graph $\Gamma$ of a directed graph $D$,
for any edge $(i^{+}, j^{-})$ between an out-vertex $i^{+}$ and an in-vertex $j^{-}$,
$i^{+}$ is a nearest neighbor of $j^{-}$, and vice versa.
The degree of $i^{+}$ ($j^{-}$) in the out-(in-)vertex set $k^{i^{+}}$ ($k^{j^{-}}$)
is the size of its nearest neighbors in the in-(out-)vertex set $|\partial i^{+}|$ ($|\partial j^{-}|$).
The degree distribution $P_{+} (k_{+})$ [$P_{-}(k_{-})$] of $\Gamma$
is the probability of randomly finding
a vertex with a degree $k_{+}$ ($k_{-}$) in the out-(in-)vertex set.
The excess degree distribution $Q_{+} (k_{+})$ [$Q_{-}(k_{-})$]  of $\Gamma$
is the probability of arriving at
a vertex with a degree $k_{+}$ ($k_{-}$) in the out-(in-)vertex set
following  a randomly chosen edge.
The arc density $c$ is defined as $c \equiv M / N$.
We simply have the equivalence
$Q_{\pm} (k_{\pm}) = k_{\pm} P_{\pm} (k_{\pm}) / c$.
When an edge $(i^{+}, j^{-})$ is removed from the original graph $\Gamma$,
we consider the residual subgraph as the cavity graph $\Gamma \backslash (i^{+}, j^{-})$.
When an out-vertex $i^{+}$ (an in-vertex $j^{-}$) is removed along with all its adjacent edges from $\Gamma$,
we consider the residual subgraph as the cavity graph $\Gamma \backslash i^{+}$ ($\Gamma \backslash j^{-}$).

Our analytical framework is based on the intuition of the Karp-Sipser algorithm
and has two components: the core percolation theory and the perfect matching of cores.
The core percolation theory
\cite{
Bauer.Golinelli-EPJB-2001,
Liu.Csoka.Zhou.Posfai-PRL-2012,
Zhao.Zhou-arXiv-2018}
is the analytical theory of the GLR procedure,
which estimates the sizes of vertices in cores and roots
as the two faces of the GLR procedure on both undirected and directed uncorrelated random graphs.
We follow the language of
the cavity method
\cite{
Mezard.Montanari-2009}
to present the core percolation theory,
in which the main results are basically marginal probabilities
calculated from the stable solutions of pertinent cavity probabilities defined on random graphs.
For the problem here on an uncorrelated random undirected bipartite graph $\Gamma$,
a set of four cavity probabilities is introduced.
Following a randomly chosen edge $(i^{+}, j^{-})$ arriving at the out-vertex $i^{+}$,
$\alpha ^{+}$ ($\beta ^{+}$)
is defined as the probability that $i^{+}$ becomes a leaf (a root)
in the GLR procedure on the cavity graph $\Gamma \backslash (i^{+}, j^{-})$;
following a randomly chosen edge $(i^{+}, j^{-})$ arriving at the in-vertex $j^{-}$,
$\alpha ^{-}$ ($\beta ^{-}$)
is defined as the probability that $j^{-}$ becomes a leaf (a root)
in the GLR procedure on $\Gamma \backslash (i^{+}, j^{-})$.
On random undirected bipartite graphs without degree-degree correlations,
with the locally tree-like structure approximation
\cite{Mezard.Montanari-2009},
we have the self-consistent equations for $\alpha ^{\pm}$ and $\beta ^{\pm}$.

\begin{eqnarray}
\alpha ^{\pm}
\label{eq:alpha}
&&
= \sum _{k_{\pm} = 1}^{\infty}
Q_{\pm} (k_{\pm}) (\beta ^{\mp}) ^{k_{\pm} - 1}, \\
\beta ^{\pm}
\label{eq:beta}
&&
= 1 - \sum _{k_{\pm} = 1}^{\infty}
Q_{\pm} (k_{\pm}) (1 - \alpha ^{\mp}) ^{k_{\pm} - 1}.
\end{eqnarray}
With the stable fixed solutions of $(\alpha ^{\pm}, \beta ^{\mp})$,
we can calculate the fractions of out- and in-vertices
in the core respectively as $n^{\pm}$,
and the fraction of all roots in the out- and in-vertex sets
as $w$.

\begin{eqnarray}
\label{eq:n}
n^{\pm}
&&
= \sum _{k_{\pm} = 2}^{\infty}
P_{\pm} (k_{\pm}) \nonumber \\
&&
\times \sum _{s = 2}^{k_{\pm}}
 \left(\begin{array}{c} k_{\pm} \\ s \end{array}\right)
(\beta ^{\mp}) ^{k_{\pm} - s}
(1 - \alpha ^{\mp} - \beta ^{\mp}) ^{s}, \\
\label{eq:w}
w
&&
= \big[ 1 - \sum _{k_{+} = 0}^{\infty}
P_{+} (k_{+}) (1 - \alpha ^{-}) ^{k_{+}} \big] \nonumber \\
&&
+ \big[ 1 - \sum _{k_{-} = 0}^{\infty}
P_{-} (k_{-}) (1 - \alpha ^{+}) ^{k_{-}} \big]
- c \alpha ^{+} \alpha ^{-}.
\end{eqnarray}
The equation for $n^{\pm}$ can be further simplified as follows.
We first move the summation $\sum _{s = 2}^{k_{\pm}}$ to $\sum _{s = 0}^{k_{\pm}}$,
then move the summation $\sum _{k_{\pm} = 2}^{\infty}$
to $\sum _{k_{\pm} = 0}^{\infty}$.
With the equivalence
$Q_{\pm} (k_{\pm}) = k_{\pm} P_{\pm} (k_{\pm}) / c$
and Eq. (\ref{eq:alpha}),
we have a concise form for $n^{\pm}$.

\begin{eqnarray}
\label{eq:n_simplified}
n^{\pm}
&&
= \sum _{k_{\pm} = 0}^{\infty}
P_{\pm} (k_{\pm})
[(1 - \alpha ^{\mp})^{k_{\pm}} 
- (\beta ^{\mp}) ^{k_{\pm}}] \nonumber \\
&&
- c \alpha ^{\pm} (1 - \alpha ^{\mp} - \beta ^{\mp}).
\end{eqnarray}
%
%
Equations (\ref{eq:alpha}), (\ref{eq:beta}), and (\ref{eq:n})
are first derived in
\cite{
Liu.Csoka.Zhou.Posfai-PRL-2012},
while Eq. (\ref{eq:w}) is our contribution here.
We further present a simple explanation of the four equations.
On a cavity graph $\Gamma \backslash i^{+}$
as $i^{+} \in \Gamma$ is a randomly chosen out-vertex,
the locally tree-like structure approximation assumes that
the states
(being leaves, roots, or trivial isolated vertices)
of $i^{+}$'s nearest neighbors or $\partial i^{+}$
are independent of each other in the GLR procedure.
The same idea also applies on $\Gamma \backslash j^{-}$ upon a randomly chosen in-vertex $j^{-}$.
We first consider the case from a cavity graph $\Gamma \backslash i^{+}$
to another cavity graph $\Gamma \backslash (i^{+}, j^{-})$
after some edges are added
in which $(i^{+}, j^{-})$ is a randomly chosen edge on $\Gamma$.
If $i^{+}$ is a leaf on $\Gamma \backslash (i^{+},  j^{-})$,
its nearest neighbors other than $j^{-}$ or simply $\partial i^{+} \backslash j^{-}$
should be all roots on $\Gamma \backslash i^{+}$.
The same logic also applies for $j^{-}$ to be a leaf
in $\Gamma \backslash (i^{+},  j^{-})$.
Thus we have Eq. (\ref{eq:alpha}).
If $i^{+}$ is a root on $\Gamma \backslash (i^{+},  j^{-})$,
there should be at least one leaf in $\partial i^{+} \backslash j^{-}$
on $\Gamma \backslash i^{+}$.
The same logic also applies for $j^{-}$ to be a root
in $\Gamma \backslash (i^{+},  j^{-})$.
Thus we have Eq. (\ref{eq:beta}).
We then consider the case from a cavity graph $\Gamma \backslash i^{+}$ to the original graph $\Gamma$
after some edges are added in which $i^{+}$ is a randomly chosen out-vertex on $\Gamma$.
If $i^{+}$ is in the core on $\Gamma$,
then among all its nearest neighbors or simply $\partial i^{+}$ on $\Gamma \backslash i^{+}$,
there should be no leaves
and also at least two vertices in the core to forbid the GLR procedure.
The same logic also applies for $j^{-}$ to be in the core on $\Gamma$.
Thus we have Eq. (\ref{eq:n}).
If $i^{+}$ is a root on $\Gamma$,
there should be at least one leaf in $\partial i^{+}$ on $\Gamma \backslash i^{+}$.
The same logic also applies for $j^{-}$ to be a root on $\Gamma$.
Yet a recounting happens
in which the contribution of an isolated edge
to a matching is counted twice.
For example, see the isolated edge $(2^{+}, 3^{-})$ in Figs. \ref{fig:model} (e) and \ref{fig:model} (f).
To calculate the probability of the formation of isolated edges,
we consider the case from a cavity graph $\Gamma \backslash (i^{+}, j^{-})$ to the original graph $\Gamma$
after the edge $(i^{+}, j^{-})$ is added
in which $(i^{+}, j^{-})$ is a randomly chosen edge on $\Gamma$.
If $(i^{+}, j^{-})$ is an isolated edge on $\Gamma$,
both $i^{+}$ and $j^{-}$ should be leaves on $\Gamma \backslash (i^{+}, j^{-})$.
Thus we have Eq. (\ref{eq:w}).

Here we explain a little more on the configurations of core and roots
on an undirected bipartite graph.
The core from the GLR procedure is well defined
\cite{
Bauer.Golinelli-EPJB-2001},
that is to say, the configuration of a core is independent of the pruning process.
Thus it is reasonable to quantify the fractions of out- and in-vertices in a core respectively as $n^{\pm}$.
Yet the configuration of the roots is dependent on the pruning process.
For example, a specific pruning process,
in which all leaves in the in-vertex set trigger GLR steps before the leaves in the out-vertex set,
simply leads to a larger size of out-vertices as roots
on an undirected bipartite graph with degree symmetry.
Yet the size of all roots of a bipartite graph, on average, is independent of the pruning process of the GLR procedure
\cite{
Bauer.Golinelli-EPJB-2001}.
This is why we calculate the fraction of roots on a whole graph as $w$
rather than distinguishing fractions of roots in the out- and in-vertex sets as $w^{\pm}$.

The second component of our analytical framework is
the perfect matching of the core structure
\cite{
Lovasz.Plummer-1986,
Karp.Sipser-IEEFoCS-1981}.
On an undirected bipartite graph $\Gamma$,
in the case of $n^{+} > n^{-}$,
the perfect matching states that the in-vertices in the core are all matched,
leading to the MM fraction of the core structure simply as $n^{-}$.
Vice versa for the case of $n^{-} > n^{+}$.

For a given random undirected bipartite graph $\Gamma$,
summing the fraction of matched edges reconstructed from the roots of the GLR procedure,
which is simply $w$,
and the estimated fraction of matched edges in the core structure,
which is just $\min \{n^{+}, n^{-}\}$,
we have the fraction of MMs on $\Gamma$.
Equivalently, we have

\begin{eqnarray}
y
\label{eq:y}
&&
=
w + \min \{n^{+}, n^{-}\}.
\end{eqnarray}
Taken together,
Eqs. (\ref{eq:alpha}), (\ref{eq:beta}), (\ref{eq:n}), (\ref{eq:w}), and (\ref{eq:y})
constitute our analytical framework of MM fractions
on random undirected bipartite graphs.

It is easy to see that,
our theory assumes a general form of degree distributions $P_{\pm} (k_{\pm})$
for a random bipartite graph.
In the case of degree symmetry,
we have $P_{+}(k) = P_{-}(k)$ for any $k$.
From Eqs. (\ref{eq:alpha}) and (\ref{eq:beta}), 
we have $\alpha ^{+} = \alpha ^{-}$ and $\beta ^{+} = \beta ^{-}$,
further leading to $n^{+} = n^{-}$ from Eq. (\ref{eq:n}).
Equation (\ref{eq:y}) can be equivalently formulated as

\begin{eqnarray}
y
\label{eq:y_symmetry}
&&
= w+  \frac {1}{2}(n^{+} + n^{-}).
\end{eqnarray}
With Eqs. (\ref{eq:n}) and (\ref{eq:w})
inserted into the above equation, we have

\begin{eqnarray}
&& 1 - y \nonumber \\
&&
\label{eq:y_symmetry_simplified}
= \frac {1}{2}\bigg \{
\big[\sum _{k_{+} = 0}^{\infty} P_{+} (k_{+})
(1 - \alpha ^{-})^{k_{+}}
+  \sum _{k_{+} = 0}^{\infty} P_{+} (k_{+})
(\beta ^{-}) ^{k_{+}}
- 1 \big] \nonumber \\
&&
+ \big[\sum _{k_{-} = 0}^{\infty} P_{-} (k_{-})
(1 - \alpha ^{+})^{k_{-}}
+  \sum _{k_{-} = 0}^{\infty} P_{-} (k_{-})
(\beta ^{+}) ^{k_{-}}
- 1 \big] \nonumber \\
&&
+ c \big[ \alpha ^{+} (1 - \beta ^{-}) + \alpha ^{-} (1 - \beta ^{+}) \big] \bigg \}.
\end{eqnarray}
We can compare Eq. (\ref{eq:y_symmetry_simplified})
with the result in
\cite{
Liu.Slotine.Barabasi-Nature-2011}.
We substitute the parameters
$w_{1}$ in Eq. (S26) as $\alpha ^{+}$,
$w_{2}$ in Eq. (S27) as  $\beta ^{+}$,
$\hat w_{1}$ in Eq. (S29) as $\alpha ^{-}$, and
$\hat w_{2}$ in Eq. (S30) as $\beta ^{-}$.
The MDNS fraction $n_{D}$ of Eq. (S37)
simply reduces to Eq. (\ref{eq:y_symmetry_simplified}).
Thus our theory retrieves the estimation of MDNS sizes on random directed graphs with degree symmetry
based on the cavity method at zero temperature limit.
At the zero temperature limit for a finite-temperature cavity method for a physical system, 
the inverse temperature $\beta \equiv 1 / T$ with $T$ as the temperature
is assumed as $\beta \rightarrow \infty$.
In this limit case, only the ground-state solutions
(in our case, the MMs with the maximum fraction)
contribute to the physical system.
The cavity messages on random graph ensembles
thus can be coarse-grained
and alternatively denoted with the cavity probabilities
describing the distribution of these coarse-grained values.
Correspondingly, the self-consistent equations
and the ground-state energy with the cavity messages
all reduce to those forms with the cavity probabilities,
along with the topological property (degree distribution) of random graphs as input.
An intuitive understanding of this correspondence of results here and in
\cite{
Liu.Slotine.Barabasi-Nature-2011}
is that our analytical approach
as a combination of the core percolation theory and the perfect matching of cores
is essentially a cavity calculation of MM sizes on random graphs directly at the zero temperature.
An implication of this correspondence
is that Eq. (S37) for $n_{D}$ in
\cite{
Liu.Slotine.Barabasi-Nature-2011}
only applies on random directed graphs with degree symmetry,
while our theory also applies on random directed graphs without degree symmetry.
On graphs with degree asymmetry,
adopting Eq. (\ref{eq:y_symmetry}) rather than Eq. (\ref{eq:y})
leads to an overestimation of $y$ by $|n^{+} - n^{-}| / 2$.

\section{Results}
\label{sec:results}

\begin{figure*}
\begin{center}
 \includegraphics[width = 0.83 \linewidth]{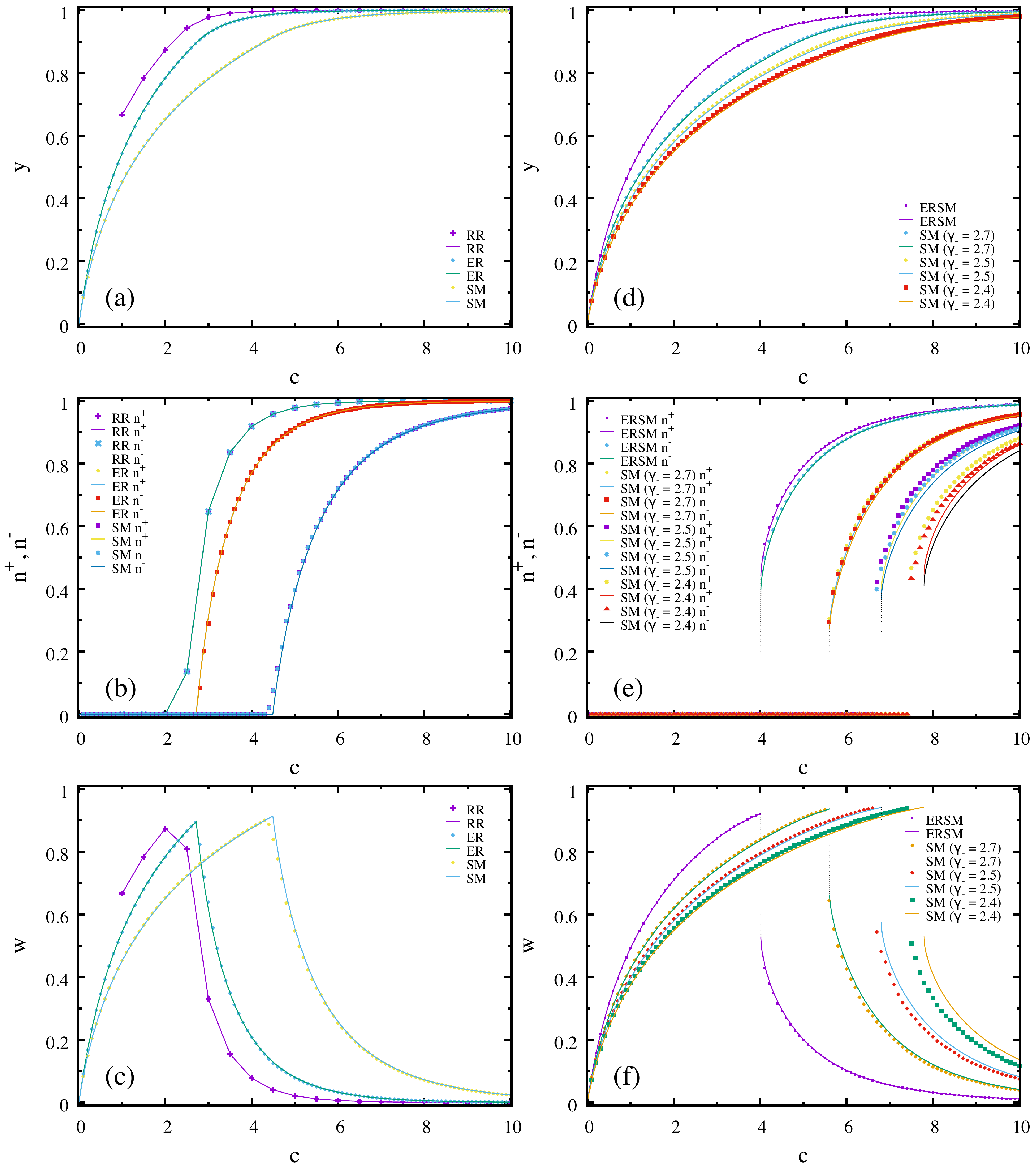} 
\end{center}
\caption{
 \label{fig:glrd_rand}
The relative sizes of MM $y$,
the fractions of out- and in-vertices in cores $n^{\pm}$,
and the fraction of roots $w$ on random directed graphs.
(a) - (c) show results on
the directed random regular (RR) graphs,
directed Erd\"{o}s-R\'{e}nyi (ER) random graphs,
and directed asymptotical scale-free networks generated with the static model (SM)
with degree exponents $\gamma _{+} = \gamma _{-} = 3.0$,
which are listed from left to right, respectively.
(d) - (f) show results on
the random directed graphs with Poissonian out-degree distribution
and power-law in-degree distribution with an exponent $\gamma _{-} = 3.0$ generated with the static model (ERSM) and
the directed asymptotical scale-free networks generated with the static model (SM)
with $\gamma _{+} = 3.0$ and $\gamma _{-} = 2.7, 2.5, 2.4$,
which are listed from left to right, respectively.
(a) and (d) show results of $y$.
(b) and (e) show results of $n^{\pm}$.
(c) and (f) show results of $w$.
Each sign is for the simulation result on a single graph instance with a vertex size $N = 10^5$.
Each solid line is for the analytical result on infinitely large random graphs.
Each dotted line in (e) and (f) is for a discontinuity in the analytical result.}
\end{figure*}
\begin{figure*}
\begin{center}
 \includegraphics[width = 0.99 \linewidth]{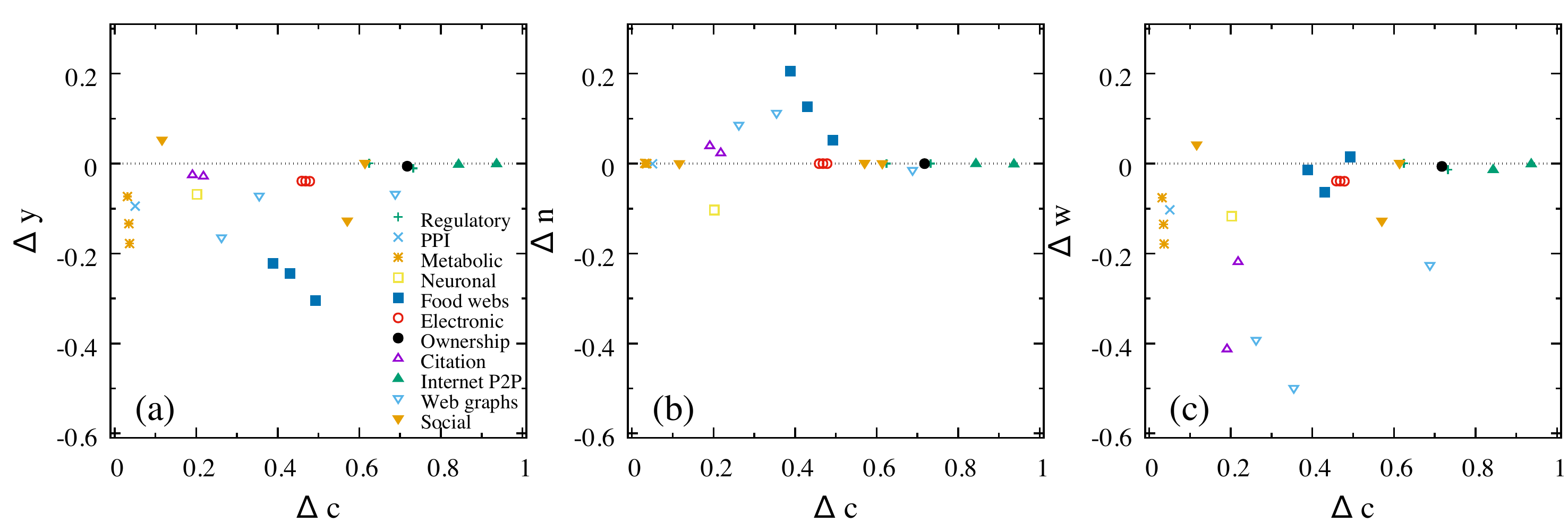} 
\end{center}
\caption{
 \label{fig:glrd_real}
The differences of
the MM sizes $\Delta y$,
the core asymmetries $\Delta n$, and
the root sizes $\Delta w$
between simulation and analytical theory
on $24$ real-world network instances.
$\Delta y$, $\Delta n$, and $\Delta w$
are shown against the degree asymmetry $\Delta c$
in (a), (b), and (c), respectively.}
\end{figure*}
%
%

We test the simulation (the GLR procedure and the Karp-Sipser algorithm)
and our analytical framework on some model random directed graphs.
The details of graph construction and simplified equations of our theory are left in Appendix A.
First, we consider the
random directed graphs with symmetric out- and in-degree distributions.
Examples are
the directed Erd\"{o}s-R\'{e}nyi random graphs
\cite{
Erdos.Renyi-PublMath-1959,
Erdos.Renyi-Hungary-1960},
the directed random regular graphs,
and the directed scale-free networks
\cite{
Barabasi.Albert-Science-1999}
generated with the static model
\cite{
Goh.Kahng.Kim-PRL-2001,
Catanzaro.PastorSatorras-EPJB-2009,
Lee.Goh.Kahng.Kim-EPJB-2006}
with the same out-degree exponent $\gamma ^{+}$ and in-degree exponent $\gamma ^{-}$.
See the results of $y$, $n^{\pm}$, and $w$ in Figs. \ref{fig:glrd_rand} (a) to \ref{fig:glrd_rand} (c).
We then consider two cases of random directed graphs with degree asymmetry.
The first case involves a same form for the out- and in-degree distributions with different parameters:
the directed scale-free networks generated with the static model
with degree exponents $\gamma _{+} \neq \gamma _{-}$.
The second case involves different forms for the out- and in-degree distributions:
the directed random graphs with a Poissonian out-degree distribution
and a power-law in-degree distribution generated from the static model.
See the result in Figs. \ref{fig:glrd_rand} (d) to \ref{fig:glrd_rand} (f).
In Fig. \ref{fig:glrd_rand},
we can see that, generally speaking,
results coincide well between finite-size simulation and infinite-size analytical theory,
except in cases of power-law degree distributions with $\gamma _{-} < 3.0$ from the static model.
This tendency of result discrepancy has a root in the intrinsic degree-degree correlations
in the graph construction process.
As from the analytical results in
\cite{
Lee.Goh.Kahng.Kim-EPJB-2006},
graphs generated with the static model with a degree exponent $\gamma$
are much like uncorrelated random graphs when $\gamma > 3.0$,
while they show increasingly recognizable degree-degree correlation
with decreasing $\gamma$ when $\gamma < 3.0$.

From
\cite{
Liu.Csoka.Zhou.Posfai-PRL-2012},
on random directed graphs with degree symmetry,
the degenerate solutions of $(\alpha ^{\pm}, \beta ^{\mp})$
experience a continuous decrease at the core percolation transitions,
leading to a continuous emergence of core with nontrivial $n^{\pm}$,
while on random directed graphs with degree asymmetry,
the stable solution of $(\alpha ^{+}, \beta ^{-})$ or $(\alpha ^{-}, \beta ^{+})$
experiences a discontinuous drop at the core percolation transitions,
leading to a discontinuous appearance of core with finite $n^{\pm}$ by a gap.
In Fig. \ref{fig:glrd_rand},
we can see that $w$ follows a rise-and-fall pattern
and undergoes a continuous decrease or a sudden drop
with the same continuity of $n^{\pm}$ at the core percolation transitions.
Here we give an intuitive understanding of the pattern of $w$ and its behavior at the transition points.
Before the formation of the giant connected component on a graph,
there are mainly trees with leaves in the graph.
The GLR procedure is carried out basically based on the existing leaves in the graph,
during which more added edges lead to more leaves and more GLR steps,
thus an increasing $w$.
With more edges added into the graph after the formation of the giant connected component,
newly revealed leaves in the iterative GLR procedure play an increasingly important role,
in which a larger arc density $c$ leads to more elementray steps of the GLR procedure,
thus an ever increasing $w$,
until the formation of a core.
A core is a subgraph which the GLR procedure cannot touch.
A sudden emergence of core $n^{\pm}$ at the core percolation transition
simply leads to a macroscopic fraction of vertices $(n^{+} + n^{-}) / 2$ in the graph
where the GLR procedure is excluded,
thus a sudden drop of the root size $w$.
With even more edges added into the graph beyond the core percolation transition,
there is an increasing difficulty both in triggering the GLR procedure and generating new leaves.
Thus an increasing $n^{\pm}$ and a shrinking $w$
happen at the same time with a growing arc density $c$.

We further apply the simulation and our theory on real-world networks.
A description of a dataset of $24$ network instances is in Appendix B.
We focus on the effect of the degree asymmetry on the matching sizes.
For a directed network instance $D$
with degree distributions $P_{\pm}(k_{\pm})$
for its undirected bipartite representation,
we define a parameter to measure its degree asymmetry as
$\Delta c \equiv \sum _{k = 0}^{\infty} |P_{+} (k) - P_{-}(k)| / 2$.
It is easy to see that $\Delta c \in [0, 1]$,
and a larger $\Delta c$ corresponds to a larger disparity between the out- and in-degree distributions.
We also define the core asymmetry towards out-vertices in cores as $\delta n \equiv n^{+} - n^{-}$
as the difference between the sizes of the out- and in-vertex sets in a core
after the GLR procedure.
For any real-world network in the dataset,
we count from simulation the fraction of MM $y_{\rm{real}}$,
the core asymmetry $\delta n_{\rm{real}}$,
and the fraction of roots $w_{\rm{real}}$.
We also calculate $y_{\rm{theory}}$, $\delta n_{\rm{theory}}$, and $w_{\rm{theory}}$ respectively
with the analytical theory with the empirical out- and in-degree distributions of the network instance as inputs.
These analytical predictions based on empirical degree distributions
can be approximately considered as averaged results of simulation
on network instances with degree-constrained randomized wiring of arcs
\cite{
Liu.Slotine.Barabasi-Nature-2011}.
The difference
$\Delta y \equiv y_{\rm{real}} - y_{\rm{theory}}$,
$\Delta n \equiv \delta n _{\rm{real}} - \delta n _{\rm{theory}}$, and
$\Delta w \equiv w_{\rm{real}} - w_{\rm{theory}}$
can be considered as measures of
the deviation of real-world networks from their randomized versions
from the perspectives of the GLR procedure and MM sizes.
Results on the real-world network dataset are in Fig. \ref{fig:glrd_real}.
We can see that the $24$ network instances show a wide range of degree asymmetry.
For the core asymmetry difference $\Delta n$,
$16$ instances show trivial values,
seven instances show clear positive values,
and only one instance shows clear negative values.
For the root size difference $\Delta w$,
eight instances show trivial values,
only one instance shows clear positive values,
and the other $15$ instances show clear negative values.
For the MM size difference $\Delta y$,
six instances show trivial values,
only one instance shows clear positive values,
and the other $17$ instances show clear negative values.
Taken the above results in short,
we have two major observations.
First, discernible nontrivial values of $\Delta y$, $\Delta n$, and $\Delta w$
mainly appear in those instances with moderate and small degree asymmetries.
Second, 
compared to their degree-constrained randomized counterparts,
the network instances in our dataset
show a rather clear tendency towards larger core asymmetries,
smaller root sizes, and smaller MM sizes.
We only present an empirical description of the results here.
Further quantitive study on how the higher order structure in real networks
affect on the deviation of core asymmetry and roots should be continued.

\section{Conclusion}
\label{sec:conclusion}

Establishing the relation between the network structural properties and the size of driver nodes or actuators
to guide a dynamical system to any final state
is a fundamental problem in the network controllability problem,
which is a recent example of the intricate interplay
between structure and function of complex networked systems.
In this paper,
we derive a simple alternative framework
to estimate the fraction of MMs on directed systems,
thus the size of MDNSs,
a basic quantity of the network controllability problem.
Our simulation method adopts the Karp-Sipser algorithm
on the undirected bipartite representations
of the underlying directed networks.
Our analytical theory is based on the core percolation theory and the perfect matching of cores,
which works on random graphs with or without degree symmetry.
We also find that real-world network instances show
a clear tendency to larger core asymmetry and smaller root sizes and MM sizes
compared to their degree-constrained randomizations,
which is worthy for further study.
We hope that our theory contributes to clarify the fundamental role
of the network structure in the controllability and control issues
of complex connected systems.

\section{Acknowledgements}
This research is supported by
the National Natural Science Foundation of China (Grant No. 11747601)
and the Chinese Academy of Sciences Grant QYZDJ-SSW-SYS018.
J.-H.Z. is partially supported by
the Key Research Program of Frontier Sciences Chinese Academy of Sciences (Grant No. QYZDB-SSW-SYS032) and
the National Natural Science Foundation of China (Grant No. 11605288).
J.-H.Z. thanks Professor Pan Zhang (ITP-CAS) for his hospitality.

\section{Appendix A: Equations on model random graphs}

\subsection{Erd\"{o}s-R\'{e}nyi random graphs}

For directed Erd\"{o}s-R\'{e}nyi (ER) random graphs
\cite{
Erdos.Renyi-PublMath-1959,
Erdos.Renyi-Hungary-1960},
a graph instance $D = \{V, E\}$
with a vertex set $V$ of a size $N (= |V|)$ and an arc set $E$ of a size $M (= |E|)$
can be constructed as the following:
a null graph with $N$ vertices and no edge is initialized;
a number of $M$ arcs are established by
randomly choosing two distinct vertices, say vertices $i$ and $j$,
and connecting them with assigning a random direction with an equal probability,
say $\{i, j\}$ or $\{j, i\}$.

On directed ER random graphs with arc density $c (\equiv M / N)$,
we have the degree distributions and the excess degree distributions
of their undirected bipartite representations as

\begin{eqnarray}
P_{\pm}(k_{\pm})
&&
= e^{- c}
\frac {c^{k_{\pm}}}{k_{\pm}!}, \\
Q_{\pm}(k_{\pm})
&&
= e^{- c}
\frac {c^{k_{\pm} - 1}}{(k_{\pm} - 1)!}.
\end{eqnarray}
We have the summation rules with above equations as follows:

\begin{eqnarray}
\sum _{k_{\pm} = 0}^{\infty} P_{\pm} (k_{\pm}) x^{k_{\pm}}
&&
= e^{-c (1 - x)}, \\
\sum _{k_{\pm} = 1}^{\infty} Q_{\pm} (k_{\pm}) x^{k_{\pm} - 1}
&&
= e^{-c (1 - x).}
\end{eqnarray}
Thus we have the simplified equations as follows.

\begin{eqnarray}
\alpha ^{\pm}
&&
= e^{- c (1 - \beta ^{\mp})}, \\
\beta ^{\pm}
&&
= 1 - e ^{- c \alpha ^{\mp}}, \\
n^{\pm}
&&
= 1 - \alpha ^{\pm} - \beta ^{\pm}
- c \alpha ^{\pm} (1 - \alpha ^{\mp} - \beta ^{\mp}), \\
w
&&
= \beta ^{+} + \beta ^{-}  - c \alpha ^{+} \alpha ^{-}.
\end{eqnarray}

\subsection{Random regular graphs}

A directed random regular (RR) graph instance $D = \{V, E\}$
with a vertex set $V$ of a size $N (= |V|)$ and an arc set $E$ of a size $M (= |E|)$
can be generated from its undirected counterpart:
an undirected RR graph instance is constructed
in which each vertex is connected to an integer $K (\equiv 2 M / N)$ of randomly chosen distinct vertices;
then each edge is assigned with a random direction with an equal probability.

For the directed RR graph instances with an arc density $c (\equiv K / 2)$
as $K$ is the integer degree of the underlying undirected RR graphs,
we have the degree distributions as

\begin{eqnarray}
P_{\pm} (k_{\pm})
&&
=  \left(\begin{array}{c} K \\ k_{\pm} \end{array}\right) / 2^{K}, \\
Q_{\pm} (k_{\pm})
&&
=  \left(\begin{array}{c} K - 1 \\ k_{\pm} - 1 \end{array}\right) / 2^{K - 1}.
\end{eqnarray}
We have the summation as

\begin{eqnarray}
\sum _{k_{\pm} = 0}^{K}
P_{\pm} (k_{\pm}) x^{k_{\pm}}
&&
= (\frac {1 + x}{2})^{K}, \\
\sum _{k_{\pm} = 1}^{K}
Q_{\pm} (k_{\pm}) x^{k_{\pm} - 1}
&&
= (\frac {1 + x}{2})^{K - 1}.
\end{eqnarray}
We thus have the simplified equations as
\begin{eqnarray}
\alpha ^{\pm}
&&
= (\frac {1 + \beta ^{\mp}}{2})^{K - 1}, \\
\beta ^{\pm}
&&
= 1 - (\frac {2 - \alpha ^{\mp}}{2})^{K - 1}, \\
n^{\pm}
&&
= (\frac {2 - \alpha ^{\mp}}{2})^{K}
- (\frac {1 + \beta^{\mp}}{2})^{K} \nonumber \\
&&
- \frac {1}{2} K \alpha ^{\pm} (1 - \alpha ^{\mp} - \beta^{\mp}), \\
w
&&
= 2 - (\frac {2 - \alpha ^{-}}{2})^{K}
- (\frac {2 - \alpha ^{+}}{2})^{K}
- \frac {1}{2} K \alpha ^{+} \alpha ^{-}.
\end{eqnarray}

\subsection{Asymptotical scale-free networks}

A random directed scale-free (SF) network $D = \{V, E\}$
with $N (= |V|)$ vertices and $M (= |E|)$ arcs
follows degree distributions
$P_{+}(k_{+})  \propto k_{+}^{- \gamma _{+}} $ and $P_{-}(k_{-}) \propto k_{-}^{- \gamma _{-}}$
in which $\gamma _{+}$ and $\gamma _{-}$ are the out- and in-degree exponents, respectively.
We adopt the static model
\cite{
Goh.Kahng.Kim-PRL-2001,
Catanzaro.PastorSatorras-EPJB-2009,
Lee.Goh.Kahng.Kim-EPJB-2006}
to generate asymptotical SF networks.
We follow the procedure:
a null graph with $N$ vertices and no edge is initialized 
in which each vertex has an index $i (= 1, 2, ... N)$;
each vertex is assigned
with an out-weight
$w _{i}^{+} = \hat w _{i}^{+} \equiv i ^{- \xi _{+}}$
and an in-weight
$w _{i}^{-} = \hat w _{i}^{-} \equiv i ^{- \xi _{-}}$
as $\xi _{\pm} \equiv 1 / (\gamma _{\pm} - 1)$;
to further decouple the weights and the indices of vertices,
the out-weights and in-weights are randomly shuffled between vertices respectively,
then we have a new sequence of vertices with
an out-weight $w_{i}^{+}$ and an in-weight $w _{i}^{-}$
for each vertex $i$;
$M$ arcs are added into the null graph,
while in each step a vertex $i$ is chosen randomly from the out-vertex set with a probability proportional to its out-weight $w_{i}^{+}$,
and a vertex $j$ is chosen randomly from the in-vertex set with a probability proportional to its in-weight $w_{j}^{-}$,
then an arc $\{i, j\}$ is established if there is no $\{i, j\}$ nor $\{j, i\}$ before.

A directed SF network instance generated with the static model
with an arc density $c$,
an out-degree exponent $\gamma _{+}$,
and an in-degree exponent $\gamma _{-}$
has the degree distributions $P_{\pm}(k_{\pm})$ as

\begin{eqnarray}
P_{\pm}(k_{\pm})
& = &
\frac {1}{\xi _{\pm}}
\frac {(c (1 - \xi _{\pm}) )^{k_{\pm}}}{k_{\pm}!}
\int _{1}^{\infty}
\mathrm{d} t e^{- c (1 - \xi _{\pm}) t} t^{k_{\pm} - 1 - 1 / \xi _{\pm}} \nonumber \\
& = &
\frac {1}{\xi _{\pm}}
\frac {(c (1 - \xi _{\pm}))^{k_{\pm}}}{k_{\pm}!}
E_{- k_{\pm} + 1 + 1/\xi _{\pm}}(c (1 - \xi _{\pm})).
\end{eqnarray}
The general exponential integral function
$E_{a}(x) \equiv \int _{1}^{\infty} \mathrm{d}t e^{-xt} t^{- a}$
can be calculated with the GNU Scientific Library (GSL)
\cite{GSL}.
For large $k_{+}$ and $k_{-}$,
we have
$P_{+} (k_{+}) \propto k_{+}^{- \gamma _{+}}$ and
$P_{-} (k_{-}) \propto k_{-}^{- \gamma _{-}}$, respectively.
The excess degree distributions
$Q_{\pm} (k_{\pm})$ have the form

\begin{equation}
Q_{\pm} (k_{\pm})
=
\frac {1 - \xi _{\pm}}{\xi _{\pm}}
\frac {(c (1 - \xi _{\pm}))^{k_{\pm} - 1}}{(k_{\pm} - 1)!}
E_{- k_{\pm} + 1 + 1/\xi _{\pm}}(c (1 - \xi _{\pm})).
\end{equation}
For the summation rules, we have

\begin{eqnarray}
\sum _{k_{\pm} = 0}^{\infty} P_{\pm}(k_{\pm}) x^{k_{\pm}}
&&
= \frac {1}{\xi _{\pm}} E_{1 + \frac {1}{\xi _{\pm}}} (c (1 - \xi _{\pm}) (1 - x)), \\
\sum _{k_{\pm} = 1}^{\infty} Q_{{\pm}}(k_{\pm}) x^{k_{\pm} - 1}
&&
= \frac {1 - \xi _{\pm}}{\xi _{\pm}} E_{\frac {1}{\xi _{\pm}}} (c (1 - \xi _{\pm}) (1 - x)).
\end{eqnarray}
We thus have the simplified equations as

\begin{eqnarray}
\alpha ^{\pm}
&&
= \frac {1 - \xi _{\pm}}{\xi _{\pm}}
E _{\frac {1}{\xi _{\pm}}} (c (1 - \xi _{\pm}) (1 - \beta ^{\mp})), \\
\beta ^{\pm}
&&
= 1 - \frac {1 - \xi _{\pm}}{\xi _{\pm}}
E _{\frac {1}{\xi _{\pm}}} (c (1 - \xi_{\pm}) \alpha ^{\mp}), \\
n^{\pm}
&&
=
\frac {1}{\xi _{\pm}} E _{1 + \frac {1}{\xi _{\pm}}} (c (1 - \xi _{\pm}) \alpha ^{\mp}) \nonumber \\
&&
- \frac {1}{\xi _{\pm}} E _{1 + \frac {1}{\xi _{\pm}}} (c (1 - \xi _{\pm}) (1 - \beta ^{\mp})) \nonumber \\
&&
- c \alpha ^{\pm} (1 - \alpha ^{\mp} - \beta ^{\mp}), \\
w
&&
= 
2 - \frac {1}{\xi _{+}} E _{1 + \frac {1}{\xi _{+}}} (c (1 - \xi _{+}) \alpha ^{-}) \nonumber \\
&&
- \frac {1}{\xi _{-}} E _{1 + \frac {1}{\xi _{-}}} (c (1 - \xi _{-}) \alpha ^{+})
- c \alpha ^{+} \alpha ^{-}.
\end{eqnarray}

\subsection{Random graphs with different forms of out- and in-degree distributions}

We can construct random directed graphs
with a Poissionian out-degree distribution
and a power-law in-degree distribution generated from the static model
with an in-degree exponent $\gamma _{-}$.
In the static model, we have
$\xi _{-} \equiv 1 / (\gamma _{-} - 1)$.

With equations in the previous subsections and the arc density $c$,
we have the following simplified equations.

\begin{eqnarray}
\alpha ^{+}
&&
= e^{- c (1 - \beta ^{-})}, \\
\alpha ^{-}
&&
= \frac {1 - \xi _{-}}{\xi _{-}}
E _{\frac {1}{\xi _{-}}} (c (1 - \xi _{-}) (1 - \beta ^{+})), \\
\beta ^{+}
&&
= 1 - e ^{- c \alpha ^{-}}, \\
\beta ^{-}
&&
= 1 - \frac {1 - \xi _{-}}{\xi _{-}}
E _{\frac {1}{\xi _{-}}} (c (1 - \xi_{-}) \alpha ^{+}),\\
n^{+}
&&
= 1 - \alpha ^{+} - \beta ^{+}
- c \alpha ^{+} (1 - \alpha ^{-} - \beta ^{-}), \\
n^{-}
&&
= \frac {1}{\xi _{-}} E _{1 + \frac {1}{\xi _{-}}} (c (1 - \xi _{-}) \alpha ^{+}) \nonumber \\
&&
- \frac {1}{\xi _{-}} E _{1 + \frac {1}{\xi _{-}}} (c (1 - \xi _{-}) (1 - \beta ^{+})) \nonumber \\
&&
- c \alpha ^{-} (1 - \alpha ^{+} - \beta ^{+}), \\
w
&&
= \beta ^{+}
+ 1 - \frac {1}{\xi _{-}} E _{1 + \frac {1}{\xi _{-}}} (c (1 - \xi _{-}) \alpha ^{+})
- c \alpha ^{+} \alpha ^{-}.
\end{eqnarray}

\section{Appendix B: Description of the real network dataset}

We list some information about the real-world network dataset
we use in the main text
in table \ref{tab:real_networks}.
A major part of large network instances is from the collections in
\cite{SNAP}.
To consider the skeleton of the interaction topology among the constituents in the networked systems,
we remove self-loops (self-interaction of a constituent)
and merge multi-edges (multiple interactions with the same direction between two constituents)
in the network instances of the dataset.

\begin{table*}[htbp]
\caption{
 \label{tab:real_networks}
A list of $24$ real-world directed network instances.
For each network,
we show its type and name,
a brief description,
and its size of vertices ($N$) and arcs ($M$).}
\begin{center}
\begin{longtable}{llrr}
\hline
\hline
  Type and Name 
  & Description
  & N
  & M \\
\hline
Regulatory networks \\
 {\em E. coli}  \cite{Mangan.Alon-PNAS-2003}
 &  Transcriptional regulatory network of {\em E. coli}.
 & $418$
 & $519$ \\
 {\em S. cerevisiae} \cite{Alon.etal-Science-2002}
 & Transcriptional regulatory network of {\em S. cerevisiae}.
 & $688$
 & $1,079$ \\
 \hline
 PPI networks \\
 PPI \cite{Vinayagam.etal-ScienceSignaling-2011}
 & Protein-protein interaction network of human.
 & $6,339$
 & $34,814$ \\
\hline
Metabolic networks \\
 {\em C. elegans} \cite{Jeong.etal-Nature-2000}
 & Metabolic network of {\em C. elegans}.
 & $1,469$
 & $3,447$ \\
 {\em S. cerevisiae} \cite{Jeong.etal-Nature-2000}
 & Metabolic network of {\em S. cerevisiae}.
 & $1,511$
 & $3,833$ \\
 {\em E. coli} \cite{Jeong.etal-Nature-2000}
 & Metabolic network of {\em E. coli}.
 & $2,275$
 & $5,763$ \\
\hline
Neuronal networks \\
 {\em C. elegans} \cite{Watts.Strogatz-Nature-1998}
 & Neural network of {\em C. elegans}.
 & $297$
 & $2,345$ \\
\hline
Food weds \\
 St Marks \cite{foodweb-StMarks-1998}
 & Food web in St. Marks River Estuary.
 & $54$
 & $353$ \\
 Everglades \cite{foodweb-Everglades-2000}
 & Food web in  Everglades Graminoid Marshes. 
 & $69$
 & $911$ \\
 Florida Bay \cite{foodweb-Florida-1998}
 & Food web in Florida Bay. 
 & $128$
 & $2,106$ \\
\hline
Electronic circuits \\
 s208 \cite{Alon.etal-Science-2002}
 & Electronic sequential logic circuits.
 & $122$
 & $189$ \\
 s420 \cite{Alon.etal-Science-2002}
 & Same as above.
 & $252$
 & $399$ \\
 s838 \cite{Alon.etal-Science-2002}
 & Same as above.
 & $512$
 & $819$ \\
\hline
Ownership networks \\
 USCorp \cite{Norlen.etal-PITS14-2002}
 & Ownership network of US corporations.
 & $7,253$
 & $6,724$ \\
\hline
Citation networks \\
 cit-HepTh \cite{Leskovec.Kleinberg.Faloutsos-SIGKDD-2005}
 & Citation network in HEP-TH category of ArXiv.
 & $27,769$
 & $352,768$ \\
 cit-HepPh \cite{Leskovec.Kleinberg.Faloutsos-SIGKDD-2005}
 & Citation network in HEP-PH category of ArXiv.
 & $34,546$
 & $421,534$ \\
\hline
Internet p2p networks \\
 p2p-Gnutella04 \cite{Ripeanu.Iamnitchi.Foster-IEEEInternetComputing-2002, Leskovec.Kleinberg.Faloutsos-TransKDD-2007}
 & Gnutella peer-to-peer network from August 4, 2002. 
 & $10,876$
 & $39,994$ \\
 p2p-Gnutella31 \cite{Ripeanu.Iamnitchi.Foster-IEEEInternetComputing-2002, Leskovec.Kleinberg.Faloutsos-TransKDD-2007}
 & Gnutella peer-to-peer network from August 31, 2002.
 & $62,586$
 & $147,892$ \\
\hline
 Web graphs \\
 Notre Dame \cite{Albert.Jeong.Barabasi-Nature-1999}
 & Web graph of Notre Dame.
 & $325,729$
 & $1,469,679$ \\
 Stanford \cite{Leskovec.etal-InternetMath-2009}
 & Web graph of Stanford.edu. 
 & $281,903$
 & $2,312,497$ \\
 Google \cite{Leskovec.etal-InternetMath-2009}
 & Web graph from Google.
 & $875,713$
 & $5,105,039$ \\
\hline
Social networks \\
 WikiVote \cite{Leskovec.Huttenlocher.Kleinberg-SIGCHI-2010, Leskovec.Huttenlocher.Kleinberg-ICWWW-2010}
 & Who-vote-whom network of Wikipedia users. 
 & $7,115$
 & $103,689$ \\
 Epinions \cite{Richardson.etal-ISWC-2003}
 & Who-trust-whom network of Epinions.com users. 
 & $75,879$
 & $508,837$ \\
 Email-EuAll \cite{Leskovec.Kleinberg.Faloutsos-TransKDD-2007}
 & Email network from a EU research institution. 
 & $265,009$
 & $418,956$ \\
\hline
\hline
\end{longtable}
\end{center}
\end{table*}

\end{document}